\documentclass[aps,prd,superscriptaddress,showpacs,showkeys,floatfix,twocolumn,nofootinbib]{revtex4}

\usepackage{amsmath,amssymb,amsfonts,dcolumn,mathcomp,slashed,dsfont}
\usepackage{graphicx}
\usepackage{citesort}

\newcommand{\M}{\mathcal{M}}
\newcommand{\F}{\mathcal{F}}
\newcommand{\BR}{\mathcal{B}}
\newcommand{\Fpi}{F_\pi}

\newcommand{\disc}{{\rm disc}\,}
\newcommand{\mpn}{M_{\pi^0}}
\newcommand{\mpc}{M_\pi}

\newcommand{\mrr}{M_{\rho'}}
\newcommand{\mrrr}{M_{\rho''}}
\newcommand{\grr}{\Gamma_{\rho'}}
\newcommand{\grrr}{\Gamma_{\rho''}}
\newcommand{\nnnl}{\nonumber\\}

\begin{document}

\title{\boldmath{$\omega\to\pi^0\gamma^*$} and \boldmath{$\phi\to\pi^0\gamma^*$} transition form factors in dispersion theory}

\author{Sebastian P. Schneider}
\email{schneider@hiskp.uni-bonn.de}
\affiliation{Helmholtz-Institut f\"ur Strahlen- und Kernphysik (Theorie) and\\
             Bethe Center for Theoretical Physics,
             Universit\"at Bonn,
             D-53115 Bonn, Germany}

\author{Bastian Kubis}
\email{kubis@hiskp.uni-bonn.de}
\affiliation{Helmholtz-Institut f\"ur Strahlen- und Kernphysik (Theorie) and\\
             Bethe Center for Theoretical Physics,
             Universit\"at Bonn,
             D-53115 Bonn, Germany}

\author{Franz Niecknig}
\email{niecknig@hiskp.uni-bonn.de}
\affiliation{Helmholtz-Institut f\"ur Strahlen- und Kernphysik (Theorie) and\\
             Bethe Center for Theoretical Physics,
             Universit\"at Bonn,
             D-53115 Bonn, Germany}

%\date{\today}

\begin{abstract}
We calculate the $\omega\to\pi^0\gamma^*$ and $\phi\to\pi^0\gamma^*$ 
electromagnetic transition form factors based on dispersion theory, relying solely on 
a previous dispersive analysis of the corresponding three-pion decays and the pion vector form factor.
We compare our findings to recent measurements of the $\omega\to\pi^0\mu^+\mu^-$ decay spectrum by the NA60 collaboration,
and strongly encourage experimental investigation of the Okubo--Zweig--Iizuka-forbidden
$\phi\to\pi^0\ell^+\ell^-$ decays in order to understand the strong deviations from vector-meson dominance
found in these transition form factors.
\end{abstract}



\maketitle

\section{Introduction}

In recent years, there has been intense renewed interest in light-meson transition form factors
due to their potential role in the theoretical determinations of the anomalous magnetic moment 
of the muon (see Ref.~\cite{JegerlehnerNyffeler} for a review).  With more and more exclusive
channels contributing to the hadronic vacuum polarization measured experimentally with unprecedented
precision, it is believed that the hadronic contribution to light-by-light scattering may soon
constitute the dominant uncertainty~\cite{BijnensPrades}.  
While a full determination of the light-by-light scattering tensor remains a formidable task,
a combination of experimental data and theoretical analyses may help to constrain one of the 
most important contributions (and one of the few that are model-independently accessible), namely
the pseudoscalar ($P=\pi^0$, $\eta$, $\eta'$) pole terms.  
Their strength is determined e.g.\ for the $\pi^0$ pole contribution by the decay $\pi^0\to\gamma^*\gamma^*$,
given in terms of the doubly-virtual form factor $F_{\pi^0\gamma^*\gamma^*}(\mpn^2,q_1^2,q_2^2)$
(see e.g.\ Ref.~\cite{JegerlehnerNyffeler} for precise definitions), where $q_{1/2}^2$ denote the two
photon virtualities.  As these doubly-virtual form factors, that are to be measured 
in the rare decays $P\to \ell^+\ell^-\ell'^+\ell'^-$ (with branching ratios of the order of $10^{-5}$),
are difficult to determine precisely in experiment, it is useful to note that they are intimately linked 
(for specific values of one of the photon virtualities) to vector-meson conversion decays:
e.g., the form factor $F_{\pi^0\gamma^*\gamma^*}(\mpn^2,q^2,M_\omega^2)$ determines the dilepton spectrum
in $\omega\to\pi^0\ell^+\ell^-$, the form factor $F_{\eta\gamma^*\gamma^*}(M_\eta^2,q^2,M_\phi^2)$
can be measured in $\phi\to\eta\ell^+\ell^-$ etc.  %\textbf{XXX address isovector background issue XXX}

The interactions of hadrons with (real and virtual) photons are often thought to be described at least
to good approximation in the picture of vector-meson dominance (VMD): the $q^2$-dependence of the form factors
above should largely be given by the propagator of a light intermediate vector meson ($\rho$, $\omega$, $\phi$),
see e.g.\ Refs.~\cite{UGM-PhysRept,Klingl,Faessler,Ivashyn}.
What is interesting about the vector-meson conversion decays is that they show a very clear deviation
from such a simple VMD picture, as has been established in the decay 
$\omega\to\pi^0\mu^+\mu^-$~\cite{LeptonG,NA60,NA60new}, and also in 
$\phi\to\eta e^+ e^-$~\cite{SND-phieta,Zdebik}.

In this article, we will analyze two such vector-meson transition form factors with the method
of dispersion relations, concentrating on $\omega\to\pi^0\gamma^*$ and 
$\phi\to\pi^0\gamma^*$ (the latter being rarer due to the implied violation of the Okubo--Zweig--Iizuka rule;
see Ref.~\cite{Pacetti} for a recent theoretical work).
One specific theoretical advantage of vector-meson conversion decays, as opposed to the pseudoscalar Dalitz 
decays, is that the isospin of the virtual photon is fixed (in the approximation that isospin is conserved).
In the cases at hand, it needs to be an iso\emph{vector} photon, hence the lowest-lying intermediate states
to contribute in a dispersion relation are $2\pi$, $4\pi$ etc.\ only, and experience with pion--pion P-wave
interactions suggests that the $2\pi$ intermediate state will already saturate the dispersion relation to a large degree.
As we will demonstrate below, a dispersive reconstruction of the $2\pi$ contribution requires two amplitudes
as input: the corresponding $V \to \pi^+\pi^-\pi^0$ decay amplitude (in the appropriate partial wave), and 
the pion (electromagnetic) vector form factor.

An analysis of the $\omega\to\pi^0\gamma^*$ transition form factor using dispersion theory has already been
performed decades ago~\cite{Koepp} (although phenomenologically the focus of that work lay more on the 
$e^+e^-\to\omega\pi^0$ production cross section).  
The reasons to take up this subject again are mani\-fold:
we now have much more accurate experimental as well as theoretical input at our disposal, both for the 
pion vector form factor and the required pion--pion phase shifts; furthermore, we have recently performed
a dispersive analysis of the three-pion decays of both $\omega$ and $\phi$ that treats final-state interactions
between all three pions rigorously~\cite{V3pi}, which can now serve as the consistent input to the investigation
of the transition form factors.  All these ingredients of the analysis will be reviewed below.

As a final introductory remark, we note that in this article, we will confine ourselves to an analysis
of the transition form factors in the kinematical region accessible in the corresponding vector-meson \emph{decays}.
We are aiming for a precision analysis and hence
refrain from analyzing also the processes $e^+e^-\to\pi^0\omega$~\cite{SND-omegapi,CMD2-omegapi,KLOE:omegapi0gamma} 
and $e^+e^-\to\pi^0\phi$~\cite{BABAR:pi0phi}, as we expect
these to be significantly more dependent on information from the excited-resonance region.

The outline of this article is as follows.  We introduce the necessary definitions concerning kinematics
and partial-wave decomposition in Sec.~\ref{sec:VpiVkin}. We discuss the dispersion relation for the
transition form factors in Sec.~\ref{sec:VpiVDR}, including the two main elements required as input:
the pion vector form factor and the $V\to\pi^+\pi^-\pi^0$ partial-wave amplitude.  
Numerical results for form factors, decay spectra, and branching ratios are presented in Sec.~\ref{sec:numres},
before we summarize in Sec.~\ref{sec:summary}.  Technical details on a representation of the pion vector form factor
including higher resonances are relegated to an Appendix.

\section{Kinematics and partial-wave decomposition}\label{sec:VpiVkin}
We consider the decays of the lightest isoscalar vector mesons into a $\pi^0$ and a dilepton pair,
\begin{equation}
 V(p_V)\to\pi^0(p_0)\ell^+(p_{\ell^+})\ell^-(p_{\ell^-}) \,,~ V=\omega/\phi\,,~ \ell = e/\mu \,.
\end{equation}
The  $V\to\pi^0\ell^+\ell^-$ amplitude is given as~\cite{Landsberg}
\begin{equation}\label{eq:AmpMl}
 \M_{V\pi^0}=ie^2\epsilon_{\mu\nu\alpha\beta}n^\mu p_0^\nu q^\alpha \frac{f_{V\pi^0}(s)}{s}\bar u_s(p_{\ell^-})\gamma^\beta v_{s'}(p_{\ell^+})~,
\end{equation}
where $q=p_{\ell^+}+p_{\ell^-}$, $s=(p_V-p_0)^2$, $n^\mu$ is the polarization vector of the vector meson, 
and $f_{V\pi^0}(s)$ is the \emph{electromagnetic transition form factor} of the vector meson. 
We will also discuss the corresponding normalized form factor,
\begin{equation}
F_{V\pi^0}(s) = \frac{f_{V\pi^0}(s)}{f_{V\pi^0}(0)} ~.
\end{equation}
The differential decay rate in terms of this amplitude can be written according to
\begin{equation}\label{eq:Decayrate}
\frac{d\Gamma_{V\to\pi^0\ell^+\ell^-}}{ds}=\frac{2\alpha^2}{9\pi}\biggl(1+\frac{2m_\ell^2}{s}\biggr)
\frac{q_{\ell\ell}(s)q_{V\pi^0}^3(s)}{M_V^3}|f_{V\pi^0}(s)|^2,
\end{equation}
with the fine-structure constant $\alpha=e^2/4\pi$ and masses of vector meson, neutral pion, and leptons
denoted by $M_V$, $\mpn$, and $m_\ell$, respectively.  The center-of-mass momenta are given by
\begin{equation}
q_{AB}^2(s)=\frac{\lambda(M_A^2,M_B^2,s)}{4s}~,
\end{equation}
where $\lambda(x,y,z)=x^2+y^2+z^2-2(xy+yz+xz)$ is the K\"all\'en function
(and with the slight notational abuse $M_\ell \doteq m_\ell$ implied).
Radiative corrections to Eq.~\eqref{eq:Decayrate} have been calculated in Ref.~\cite{KubisSchmidt}:
they require a careful selection of kinematic cuts on the additional soft-photon radiation for the 
$e^+e^-$ final state, and are small everywhere except near threshold for $\mu^+\mu^-$, where the Coulomb pole is significant.
The above relation for the $V\to\pi^0l^+l^-$ spectrum is completely determined by $f_{V\pi^0}(s)$ aside from a 
kinematical factor that is determined by the photon propagator and phase space. Note finally that the corresponding \emph{real-photon} 
total decay rate is given by
\begin{equation}\label{eq:Gammapigamma}
\Gamma_{V\to\pi^0\gamma}=\frac{\alpha(M_V^2-\mpn^2)^3}{24 M_V^3}|f_{V\pi^0}(0)|^2 ~.
\end{equation}

In establishing a dispersion relation for the $V\to\pi^0\gamma^*$ transition form factors, 
the corresponding three-pion decays $V(p_V)\to\pi^+(p_+)\pi^-(p_-)\pi^0(p_0)$ play a central role.
We define $s=(p_V-p_0)^2$, $t=(p_V-p_+)^2$, and $u=(p_V-p_-)^2$ with
$3s_0\doteq s+t+u=M_V^2+3\mpc^2$.
We note that due to technical reasons, the decay $V\to3\pi$ is treated in the isospin limit,
with $\mpn=M_{\pi^\pm}\doteq\mpc$. The amplitude is given as
\begin{equation}\label{eq:AmpMF}
 \M_{3\pi}=i\epsilon_{\mu\nu\alpha\beta}n^\mu p_+^\nu p_-^\alpha p_0^\beta \F(s,t,u)~.
\end{equation}
Neglecting the discontinuities of F and higher partial waves,\footnote{A simplified model for additional
F-wave contributions was studied in Ref.~\cite{V3pi} and found to yield entirely negligible corrections.} 
we can decompose $\F(s,t,u)$ in terms of functions
of a single variable with only a right-hand cut~\cite{V3pi},
\begin{equation}\label{eq:reconstruction}
 \F(s,t,u)=\F(s)+\F(t)+\F(u)~.
\end{equation}
For our analysis we will require the $l=1$ partial-wave projection of $\F(s,t,u)$, which is given by
\begin{equation}
 f_1(s)=\frac{3}{4}\int_{-1}^{1} d z \big(1-z^2\big)\, \F(s,t,u)~, \label{eq:f1-angint}
\end{equation}
where $z=(t-u)/(4q_{\pi\pi}(s)q_{V\pi^0}(s))$. Note that the angular integration in Eq.~\eqref{eq:f1-angint}
is a highly nontrivial issue due to the fact that the cuts in the variables $t$ and $u$
need to be avoided; compare the discussion in Appendix~B of Ref.~\cite{V3pi}.

\section{Dispersion relation for the transition form factor}\label{sec:VpiVDR}

\begin{figure}
 \centering
\includegraphics[width= 0.75\linewidth,clip=true]{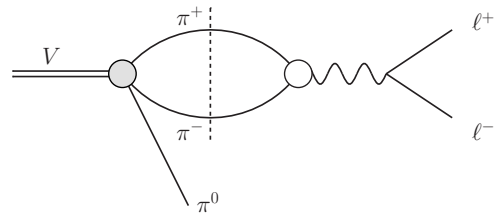} \\
\caption{Diagrammatic representation of the discontinuity of the $V\to\pi^0\ell^+\ell^-$ 
transition form factor. 
The gray circle denotes the $V\to3\pi$ amplitude, whereas the white circle represents the pion vector form factor.}
\label{fig:TFFdisc}
\end{figure}

To set up the dispersion relation for the transition form factor, 
we calculate the two-pion discontinuity of the diagram shown in Fig.~\ref{fig:TFFdisc}.
It is given as~\cite{Koepp}
\begin{equation}\label{eq:TFFdisc}
 \disc{f_{V\pi^0}(s)}=\frac{i\,q_{\pi\pi}^3(s)}{6\pi \sqrt{s}}F_\pi^{V*}(s)f_1(s)\,\theta\big(s-4\mpc^2\big)~,
\end{equation}
where $F_\pi^V(s)$ is the pion vector form factor.
Corrections to Eq.~\eqref{eq:TFFdisc} stem from heavier intermediate states of the appropriate 
quantum numbers (isospin 1 P-wave states): $4\pi$, $K\bar K$, \ldots, which are expected to be suppressed
significantly due to phase space and their higher masses. We therefore neglect these contributions in our analysis and
resort to elastic $\pi\pi$ final states only.
Given the standard assumptions on the asymptotic high-energy behavior of the pion form factor,
$F_\pi^V(s)\simeq 1/s$ (modulo logarithms),
and the $V\to 3\pi$ partial wave, $f_1(s)\simeq 1/s$~\cite{V3pi},
Eq.~\eqref{eq:TFFdisc} allows for an unsubtracted dispersion relation~\cite{Koepp}.
As our analysis, however, is confined to two-pion intermediate states and neglects any higher contributions, 
we decide to employ a once-subtracted solution of Eq.~\eqref{eq:TFFdisc} instead,
\begin{equation}\label{eq:TFFdisp}
 f_{V\pi^0}(s)= f_{V\pi^0}(0) +\frac{s}{12\pi^2}\int_{4\mpc^2}^{\infty}ds'\frac{q_{\pi\pi}^3(s')F_\pi^{V*}(s')f_1(s')}{s'^{3/2}(s'-s)} \,,
\end{equation}
in order to suppress inelastic contributions.
For the predictions of the $s$-dependence of the form factor,
we fix the subtraction constant $f_{V\pi^0}(0)$ to reproduce the 
$V\to\pi^0\gamma$ partial width according to Eq.~\eqref{eq:Gammapigamma}. 
Assuming the validity of an unsubtracted dispersion relation,
the subtraction constant and therefore the $V\to\pi^0\gamma$
partial width can be calculated by means of a sum rule,
\begin{equation}\label{eq:TFFsumrule}
f_{V\pi^0}(0) =\frac{1}{12\pi^2}\int_{4\mpc^2}^{\infty}ds'\frac{q_{\pi\pi}^3(s')}{s'^{3/2}}F_\pi^{V*}(s')f_1(s')~,
\end{equation}
which is expected to show a more problematic convergence behavior than the form-factor dispersion relation.
We will quote results for this sum rule in Sec.~\ref{sec:numres} in order to quantify
the potential role of heavier intermediate states in the transition form factor.
Nevertheless, Eq.~\eqref{eq:TFFsumrule} is a remarkable result:
as we will briefly reiterate below, in the elastic approximation the pion vector form factor is entirely given
in terms of the $\pi\pi$ P-wave phase shift $\delta_1^1(s)$, which also determines the $V\to3\pi$ partial wave up to 
a single subtraction constant that can be written as an overall normalization~\cite{V3pi}.   
This means that the ratio of branching ratios $\BR(V\to\pi^0\gamma)/\BR(V\to3\pi)$ is entirely determined
by $\delta_1^1(s)$, up to inelastic corrections.  This result is reminiscent of the relation between these
two decay modes utilized in Refs.~\cite{Lutz,Leupold} as the leading order of a Lagrangian framework
for vector mesons.

In the following we will briefly discuss the two ingredients to the dispersion integral,  
the pion vector form factor $F_\pi^V(s)$ and the $V\to3\pi$ partial-wave amplitude $f_1(s)$.

\subsection{Pion vector form factor}\label{sec:FVpi}

In the elastic approximation,
the pion vector form factor fulfills the unitarity relation
\begin{equation}\label{eq:VFF}
 \disc{F_\pi^V(s)}=2iF_\pi^V(s)\theta(s-4\mpc^2)\sin\delta(s)e^{-i\delta(s)}~,
\end{equation}
where $\delta(s)\doteq\delta_1^1(s)$ is the $\pi\pi$ P-wave phase shift. 
The solution to Eq.~\eqref{eq:VFF} is given by the Omn\`es function,
\begin{align}\label{eq:FFOmnes}
 F_\pi^V(s)=\Omega(s) =\exp\biggl\{\frac{s}{\pi}\int_{4\mpc^2}^\infty ds'\frac{\delta(s')}{s'(s'-s)}\biggr\}~,
\end{align}
normalized to $\Omega(0)=1$. 
The omission of a polynomial in $s$ multiplying the Omn\`es function relies on the absence of zeros
in the form factor, see Ref.~\cite{Anant}.
In a precision analysis of the form factor extending beyond 1~GeV, one has to account for the onset of inelasticities
(dominantly $4\pi$ intermediate states), and, as far as data extracted from $e^+e^-\to\pi^+\pi^-$ is concerned,
$\rho$--$\omega$ mixing.
As we do not have a consistent treatment of inelasticity effects in the $V\to3\pi$ partial wave $f_1(s)$
at our disposal (let alone isospin breaking), we refrain from doing so.

\begin{figure}
 \centering
\includegraphics[width= \linewidth]{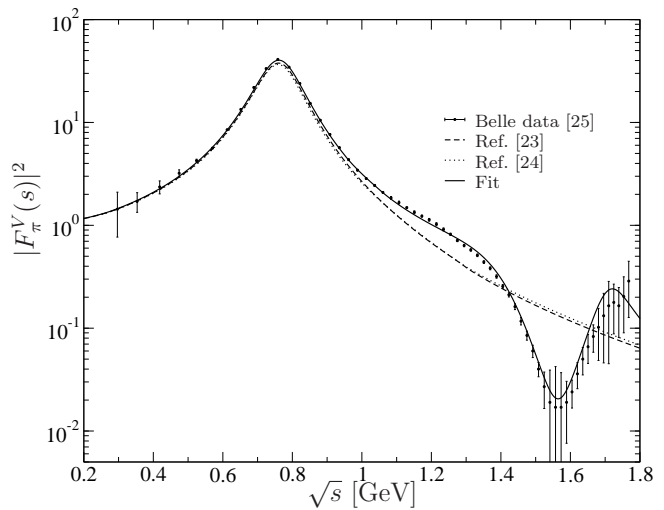} \\
\caption{Pion vector form factor fit using a phase shift incorporating elastic $\rho'$, $\rho''$ resonances 
(solid line) and solutions of the Roy equations of Refs.~\cite{Bern,Madrid} (dashed and dotted lines), 
in comparison to the experimental data of Ref.~\cite{Belle}.  For details, see main text.}
\label{fig:VFF}
\end{figure}

We use the following approach to estimate uncertainties generated by the input for the parameterization of the phase shift.
On the one hand we will
use parameterizations derived from two different solutions of the pion--pion Roy equations~\cite{Bern,Madrid},
which are valid roughly up to $1.3$~GeV. 
The pion form factor is known to excellent precision well beyond that energy
(see Refs.~\cite{Belle,BABAR,KLOE} for just the most recent experimental results),
indicating in particular contributions from the excited resonances $\rho'(1450)$ and $\rho''(1700)$.
To incorporate these higher resonance states we use the phenomenological form factor suggested in Ref.~\cite{Roig}
(and briefly summarized in Appendix~\ref{app:resFF}), which we fit to the experimental data of Ref.~\cite{Belle},
extract the corresponding phase, and match it smoothly to the phase-shift solution of Ref.~\cite{Bern} below 1~GeV.
The aforementioned procedure treats the higher resonances $\rho'$ and $\rho''$ as purely elastic,
which they clearly are not (compare the more sophisticated form factor representation of Ref.~\cite{HanhartFF});
we merely use the phase thus obtained as an indicator for uncertainties generated  in the 
energy range between roughly 1.3~GeV and 1.9~GeV.

As the Omn\`es representation requires the $\pi\pi$ P-wave phase shift up to infinity,
we have to make assumptions about its asymptotic behavior.
We choose to smoothly guide $\delta(s)$ to $\pi$, 
so that we guarantee the correct asymptotic behavior of $F_\pi^V(s)\to s^{-1}$ for $s\to\infty$. 
The point beyond which the asymptotic behavior sets in is chosen to be $\Lambda_\delta=1.3$~GeV for the Roy-equation 
analyses~\cite{Bern,Madrid},  and $\Lambda_\delta=1.9$~GeV for the phase derived from the form factor. 
The resulting form factors corresponding to the different phases
are shown in Fig.~\ref{fig:VFF}, compared to the data of Ref.~\cite{Belle}.

\subsection{\boldmath{$V\to3\pi$} partial-wave amplitude}

We only very briefly summarize the basics of the dispersive analysis of the $V\to3\pi$ decay amplitudes
and refer to Ref.~\cite{V3pi} for further details.
The partial-wave amplitude $f_1(s)$ fulfills the unitarity relation
\begin{align}\label{eq:unrel}
 \disc f_1(s) &= \disc\F(s) \\
	      &= 2i\,\bigl(\F(s)+\hat\F(s)\bigr)\theta(s-4\mpc^2)\sin\delta(s)e^{-i\delta(s)}\nonumber~,
\end{align}
where the inhomogeneity $\hat\F(s)$ is given by angular averages over $\F$ according to
\begin{align}\label{eq:inhomogen}
 \hat\F(s)&=3\langle(1-z^2)\F\rangle(s)~,\\
\langle z^n f\rangle(s)&=\frac{1}{2}\int_{-1}^1dz\, z^n f\biggl(\frac{3s_0-s}{2}+2q_{\pi\pi}(s)q_{V\pi^0}(s)z\biggr)\,.\nonumber
\end{align}
The function $\hat\F(s)$ contains the left-hand-cut contributions due to crossed-channel singularities.
In the case at hand, the left-hand cut overlaps with the right-hand one, as for $M_V>3\mpc$,
$s$, $t$, and $u$ can be simultaneously larger than $4\mpc^2$, which they are inside the physical decay region.
In this sense, there is no meaningful way from the point of view of dispersion theory to neglect the left-hand cut here.
The solution of Eq.~\eqref{eq:unrel} is given by a once-subtracted dispersion relation,
\begin{equation}\label{eq:inteeqn}
\F(s)=\Omega(s)\biggr\{a + \frac{s}{\pi}\int_{4\mpc^2}^{\infty}\frac{ds'}{s'}\frac{\sin\delta(s')\hat\F(s')}{|\Omega(s')|(s'-s)}\biggr\}~,
\end{equation}
where the subtraction constant $a$ serves as an overall normalization and is adjusted to reproduce the $V\to3\pi$ partial width. 

\begin{figure}[t]
 \centering
\includegraphics[width= 0.99\linewidth]{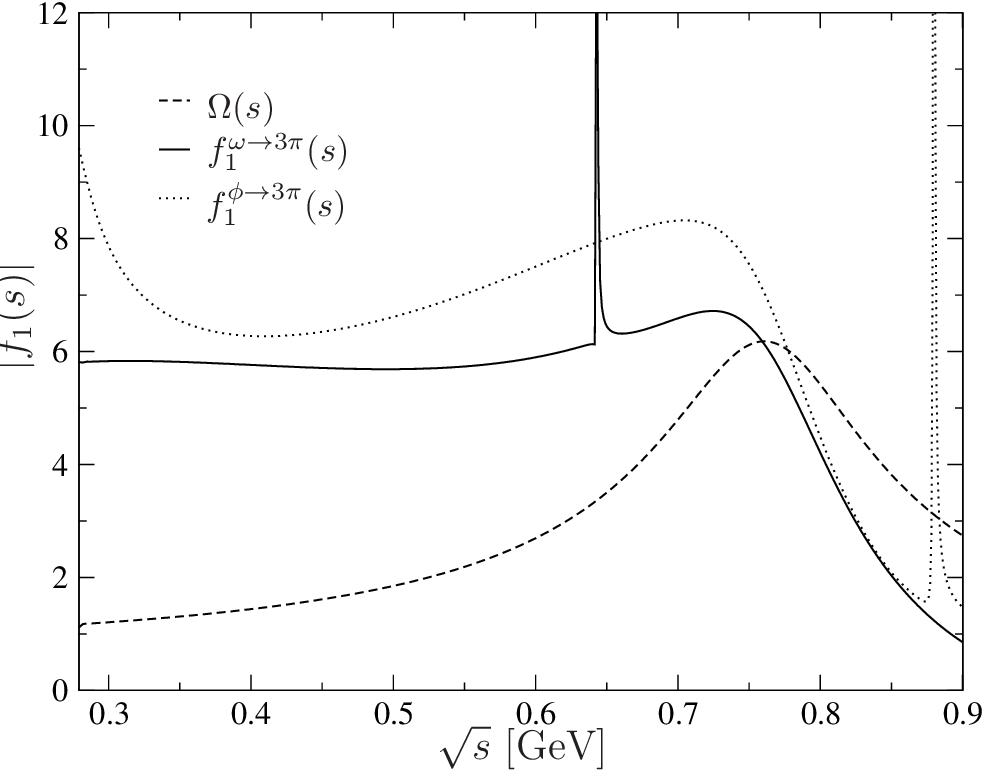} \\[2mm] \hspace*{-2mm}
\includegraphics[width= \linewidth]{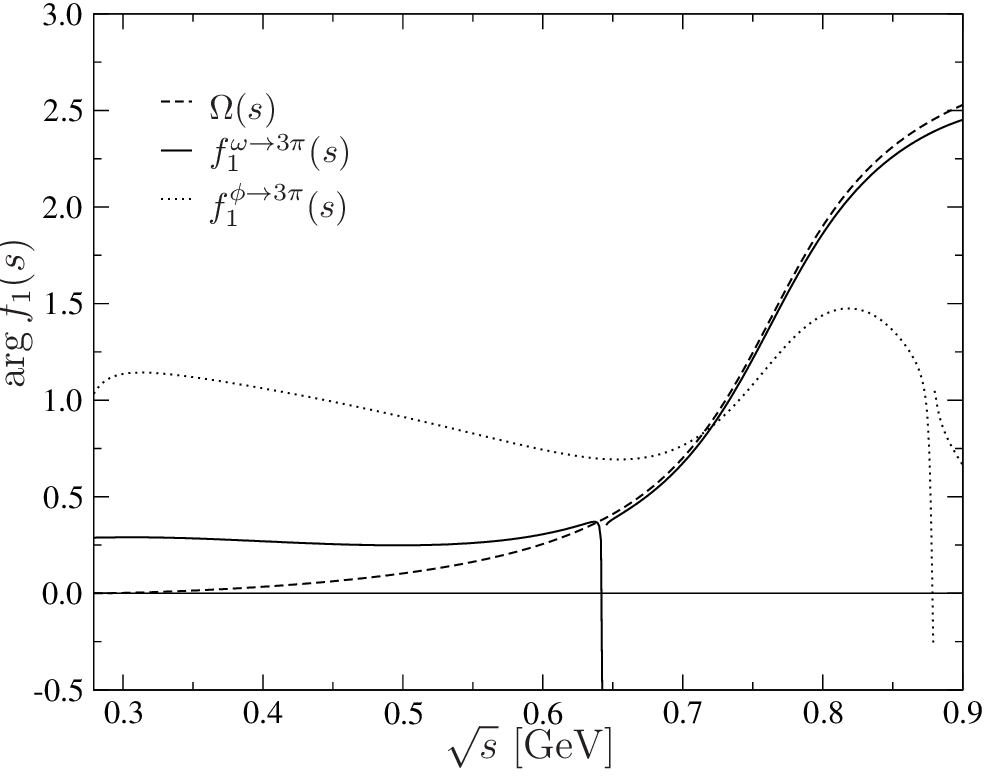} 
\caption{Modulus (upper panel) and phase (lower panel) of the P partial wave $f_1(s)$, based on Ref.~\cite{V3pi} both for 
$\omega\to3\pi$ (solid line) and $\phi\to3\pi$ (dotted line), in comparison to the Omn\`es function (dashed line). 
We refrain from devising error bands and fix the input for the phase according to 
Ref.~\cite{Bern} and the integral cutoff in Eq.~\eqref{eq:inteeqn} to $\Lambda=2.5$~GeV. 
The normalization constant $a$ is set to 1.}
\label{fig:f1}
\end{figure}
\begin{figure}
 \centering
\includegraphics[width= 0.75\linewidth,clip=true]{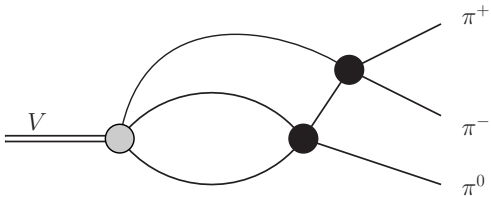} \\
\caption{Two-loop diagram contributing to the $V\to3\pi$ decay amplitude,
which has a singular discontinuity at the pseudo\-threshold $s=(M_V-M_\pi)^2$ and leads to a nonvanishing phase/
imaginary part of the corresponding partial wave $f_1(s)$ at threshold $s=4\mpc^2$.}
\label{fig:disc3pi}
\end{figure}

As Ref.~\cite{V3pi} does not explicitly show the partial-wave amplitude $f_1(s)=\F(s)+\hat\F(s)$ that plays a central role 
in the present investigation, we display its modulus and phase both for $\omega\to3\pi$ and $\phi\to3\pi$,
as derived from the numerical results in Ref.~\cite{V3pi}, compared to the Omn\`es function
(whose phase of course is just $\delta(s)$) in Fig.~\ref{fig:f1}. We note that the partial waves bear very little similarity
to the Omn\`es function: there is a strong enhancement 
in the threshold region below the $\rho$ resonance, a large part of which can be thought of as the partial-wave-projected
$t$- and $u$-channel $\rho$ exchanges in a VMD picture~\cite{Koepp}.  
Furthermore, we note that the phase of $f_1(s)$ also does not follow $\delta(s)$:
Watson's theorem does not hold due to three-pion-cut effects, see Fig.~\ref{fig:disc3pi},
which in particular allow for a nonvanishing imaginary part of $f_1(s)$ already at $\pi\pi$ threshold.\footnote{Note
that this complication does not occur in Ref.~\cite{Koepp} due to the approximation of the left-hand-cut contributions
by $\rho$ poles only, neglecting the effects of the two-pion cut starting already at $t,\,u=4\mpc^2$.  As only the 
transition form factor of the $\omega$ and not the one of the $\phi$ is considered in Ref.~\cite{Koepp}, the 
$\rho$ poles in $t$-/$u$-channel lie outside the integration range that affects the partial-wave projection.}
Finally, $f_1(s)$ shows singular behavior at the pseudothreshold $s=(M_V-\mpc)^2$, stemming from the 
diagrammatic topology shown in Fig.~\ref{fig:disc3pi}; as discussed in Refs.~\cite{V3pi,GKR}, 
these singularities in the discontinuity do \emph{not} translate into singular behavior of the form factor
itself when evaluated at the upper rim of the unitarity cut.  In particular, the irregular phase in the 
vicinity of the pseudothreshold is an artifact as a consequence of the different divergences of real and imaginary parts
from below and above, and has no physical significance.

Returning to the dispersive representation of $\F(s)$, 
it was shown in Ref.~\cite{V3pi} 
that by oversubtracting the integral equation~\eqref{eq:inteeqn} for $\phi\to3\pi$,
\begin{equation}\label{eq:oversub}
\F(s)=\Omega(s)\biggr\{a + b s + \frac{s^2}{\pi}\int_{4\mpc^2}^{\infty}\frac{ds'}{s'^2}\frac{\sin\delta(s')\hat\F(s')}{|\Omega(s')|(s'-s)}\biggr\}~,
\end{equation}
and adjusting the additional subtraction constant $b$, we were able to achieve a perfect fit of the 
$\phi\to3\pi$ Dalitz plot~\cite{KLOEphi}. 
Since the fitted value for $b$ differs from a sum rule, as suggested
by demanding the representations~\eqref{eq:inteeqn} and \eqref{eq:oversub} to be equal,
Eq.~\eqref{eq:oversub} does not satisfy the 
high-energy behavior for the partial-wave amplitude $f_1(s)$, which
therefore tends asymptotically towards a constant instead of $s^{-1}$; consequently, the
integral~\eqref{eq:TFFsumrule} does not converge, and we will not evaluate the sum rule for 
$f_{\phi\pi^0}(0)$ for the twice-subtracted solution of $f_1(s)$.

We will in general stabilize the high-energy behavior of our dispersion integrals 
by manually leading $\F(s)$ to $\Lambda^2\F(\Lambda^2)/s$ 
beyond a certain cutoff $\Lambda^2$. There is 
no obvious prescription as to when exactly the amplitude or the form factor should show this asymptotic behavior;
we choose the point up to where we have adjusted our form factor representation to data, that is $\Lambda=1.8$~GeV,
and incorporate a variation of the cutoff up to $\Lambda=2.5$~GeV in our error considerations. 
This prescription assures that we have a precision representation for the amplitude in the low-energy regime as well 
as the correct high-energy behavior. 
By varying the cutoff we assure that the intermediate-energy regime is sufficiently 
suppressed so as not to taint our numerical results, which we present in the following section.

\section{Numerical results}\label{sec:numres}

For the numerical evaluation,
we use the different parameterizations of the phase shift described in Sec.~\ref{sec:FVpi};
the same parameterization is always used consistently for both pion form factor and $V\to3\pi$ partial wave.
We vary the cutoff of the dispersion integrals in Eqs.~\eqref{eq:TFFdisp}, \eqref{eq:TFFsumrule}, and \eqref{eq:inteeqn} 
(beyond which the assumed asymptotic behavior is enforced by hand)
between $\Lambda=1.8$~GeV and $\Lambda=2.5$~GeV as detailed above.  We note that it does not make 
sense to vary the cutoff of the aforementioned integrals individually: 
the uncertainties in our treatment are related to our lack of knowledge concerning final-state interactions in the intermediate-energy range, and thus  apply equally to all considered dispersion integrals. 
The subtraction constants of the $V\to3\pi$ amplitudes are fixed by the 
total widths $\Gamma_\omega=8.49\pm0.08$~MeV and $\Gamma_\phi=4.26\pm0.04$~MeV together with
the $V\to3\pi$ branching ratios
$\BR^{\rm exp}(\omega\to3\pi) = 0.892 \pm 0.007$,
$\BR^{\rm exp}(\phi\to3\pi)= 0.153\pm0.003$,
the uncertainties of which we will always neglect in the following. 
It turns out that for all of our evaluations of the \emph{once-subtracted} 
dispersion relation in Eq.~\eqref{eq:TFFdisp}, a variation of the phase between the parameterization of Ref.~\cite{Madrid} 
and the one derived from the form factor spectrum along with an integral cutoff of $\Lambda=1.8$~GeV gives rise to an 
enveloping uncertainty band. 

For the following $V\to\pi^0\gamma$ branching ratios determined from Eq.~\eqref{eq:Gammapigamma} and the 
\emph{unsubtracted} dispersion relation \eqref{eq:TFFsumrule}, the parameterization of Ref.~\cite{Bern} and the one 
from the form factor spectrum together with an integral cutoff of $\Lambda=2.5$~GeV give rise to limiting values. 
We find
\begin{align}
\BR(\omega\to\pi^0\gamma)  &= (7.48 \ldots 7.75) \times 10^{-2} ~, \nnnl
\BR(\phi\to\pi^0\gamma)  &= (1.28\ldots 1.37) \times 10^{-3} ~,
\end{align}
which is to be checked against the experimental averages
$\BR^{\rm exp}(\omega\to\pi^0\gamma) = (8.28\pm0.28) \times 10^{-2}$, 
$\BR^{\rm exp}(\phi\to\pi^0\gamma) = (1.27\pm0.06) \times 10^{-3}$~\cite{PDG}.
We observe that the $\phi\to\pi^0\gamma$ partial width compares favorably to experiment, 
whereas the result for $\omega\to\pi^0\gamma$ turns out to be slightly too low;
even then, the $2\pi$ intermediate state seems to saturate more than 90\% of the sum rule for this partial width.
We note that the most precise individual measurement of $\BR(\omega\to\pi^0\gamma)$
actually determines the ratio of branching ratios
$
{\BR^{\rm exp}(\omega\to\pi^0\gamma)}/{\BR^{\rm exp}(\omega\to3\pi)}= (8.97\pm0.16) \times 10^{-2} 
$~\cite{KLOE:omegapi0gamma},
which is precisely the ratio we argued in Sec.~\ref{sec:VpiVDR} to be a pure prediction due to the $\pi\pi$ P-wave
phase shift, independent of any subtraction constant; for this quantity, our numerical result amounts to
\begin{equation}
\frac{\BR(\omega\to\pi^0\gamma)}{\BR(\omega\to3\pi)}= (8.39\ldots8.69) \times 10^{-2} ~,
\end{equation}
hence suggesting a saturation of the sum rule even at the 95\% level.

We stress, however, that due to the slow convergence behavior of the integrand in Eq.~\eqref{eq:TFFsumrule},
we do not consider the sum-rule results to be extremely reliable: they depend rather strongly on the assumed 
intermediate and high-energy behavior of the $\pi\pi$ phase shift. 
For example, using a cutoff of $\Lambda = 1.8$~GeV in the dispersion integral~\eqref{eq:TFFsumrule}
beyond which the asymptotic fall-off is enforced by hand, we find that this asymptotic region
$s>\Lambda^2$ still yields a 10\% correction to the $\omega\to\pi^0\gamma$ branching ratio.
We therefore rather take these as benchmark values to test the accuracy of the approximation of using only
two-pion intermediate states in the dispersion relation: we expect this to work \emph{better} 
in the description of the $s$-dependence of the transition form factor,
in which we choose the subtraction constant in Eq.~\eqref{eq:TFFdisp} fixed to the 
experimental values of the $V\to\pi^0\gamma$ partial widths. 
The errors on these values contribute a large part to the uncertainty of the transition form factor 
and the differential $V\to\pi^0\ell^+\ell^-$ decay width, which we will present in the following.

\subsection{\boldmath{$\omega\to\pi^0\ell^+\ell^-$}}

\begin{figure*}[t]
\centering
\centerline{
\parbox{0.51\linewidth}{\includegraphics[clip=true,height=11.2cm]{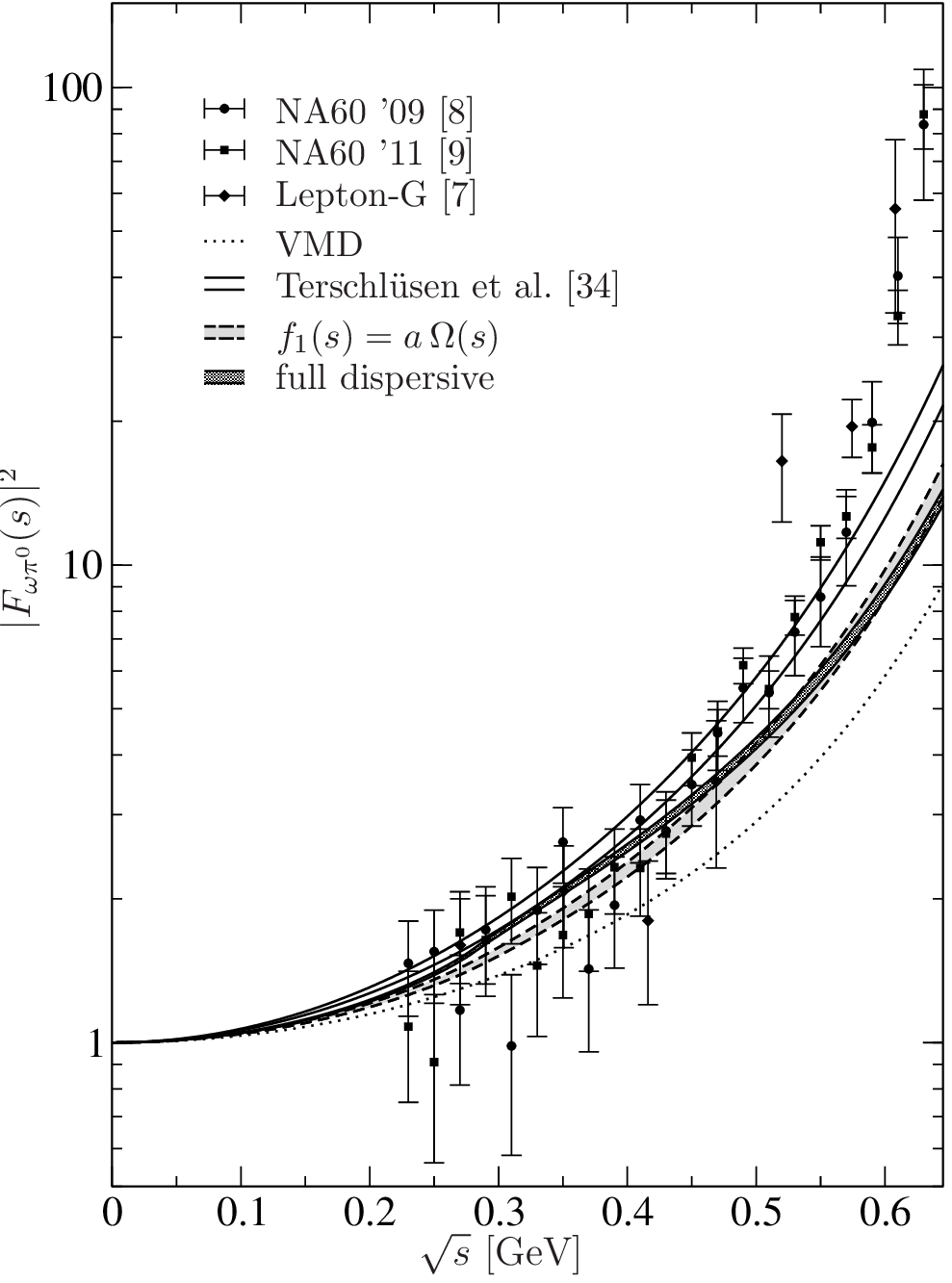}} \hfill 
\parbox{0.49\linewidth}{\vspace*{-2mm}\includegraphics[clip=true,height=5.52cm]{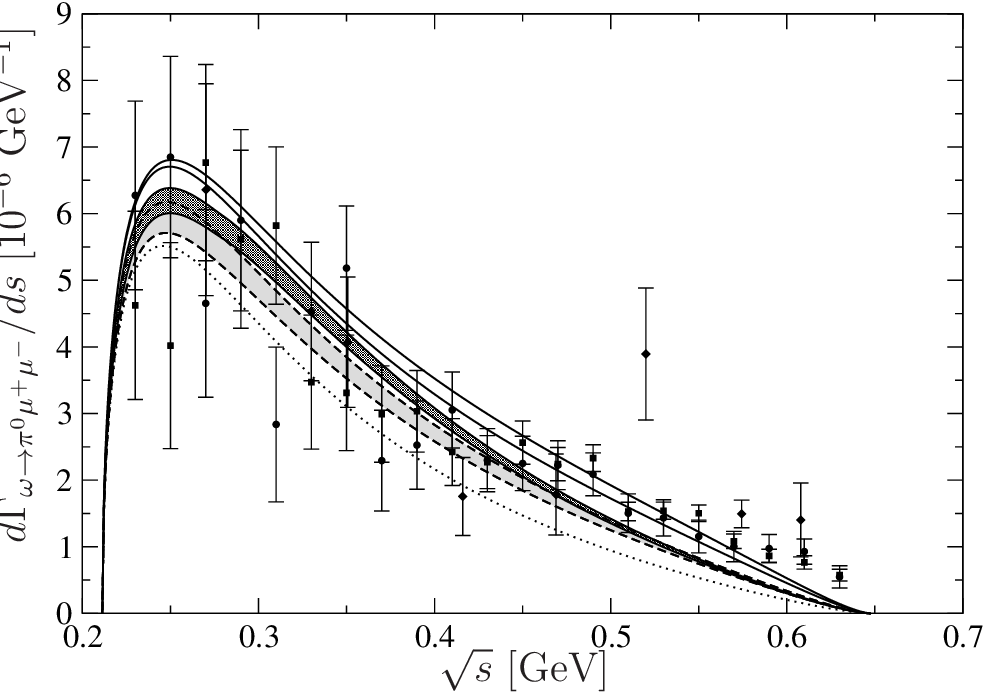} \\ \hspace*{-3.5mm}
                        \includegraphics[clip=true,height=5.65cm]{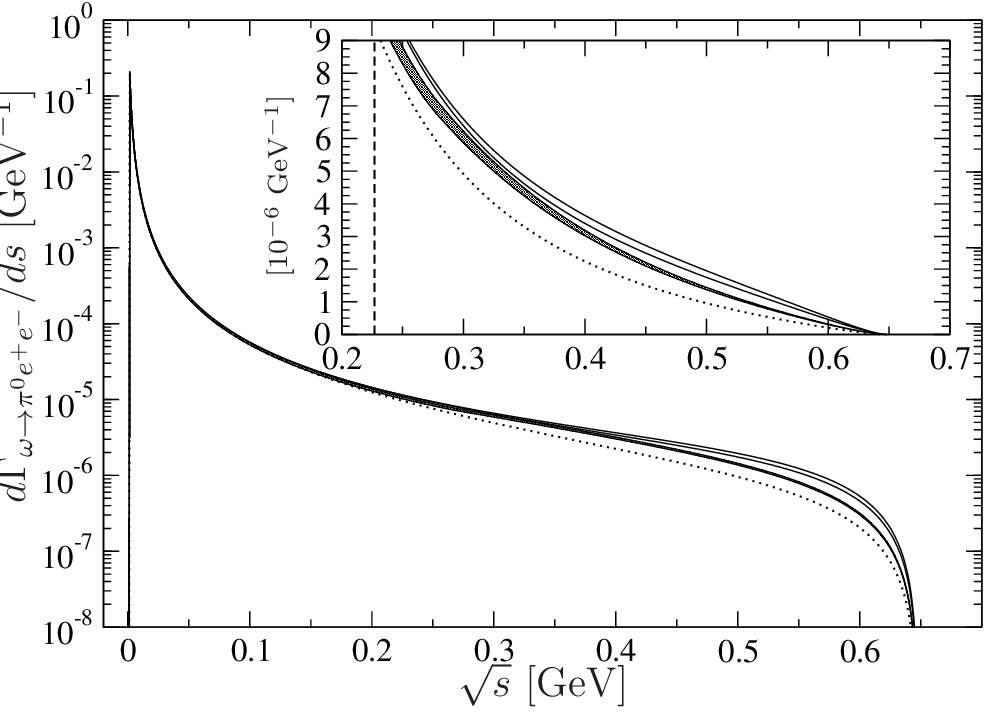}}
}
\vspace*{-0.1cm}
\caption{Left panel: Numerical results for the $\omega\to\pi^0\gamma^*$ transition form factor. 
Top, right panel: Differential $\omega\to\pi^0\mu^+\mu^-$ decay width. 
Bottom, right panel: Differential $\omega\to\pi^0e^+e^-$ decay width. 
Data for the transition form factor and the differential $\omega\to\pi^0\mu^+\mu^-$ width is taken 
from Refs.~\cite{NA60,NA60new,LeptonG} (we have not included the data set from Ref.~\cite{CMD2epem} due to its fairly low statistics). 
We show pure VMD (dotted line), 
the results of a chiral Lagrangian treatment with explicit vector mesons~\cite{Terschlusen1} 
(white shaded curve with solid borders), and the dispersive solution for $f_1(s)=a\,\Omega(s)$ 
(gray shaded curve with dashed borders) as well as the full dispersive solution 
(black hatched curve with solid borders). 
For $\omega\to\pi^0e^+e^-$ we do not display the pure Omn\`es solution, since it is 
virtually indistinguishable from the full dispersive result due to the strongly dominating 
kinematical factor in Eq.~\eqref{eq:Decayrate}. The inset magnifies the region above the two-muon threshold 
(vertical dashed line).}
\label{fig:TFF}
\end{figure*}

We start off by considering $V=\omega$. 
In Fig.~\ref{fig:TFF} we compare the absolute value squared of the $\omega\to\pi^0\gamma^*$ transition form factor (left panel) 
and the differential 
$\omega\to\pi^0\ell^+\ell^-$ decay widths (right panel) calculated in our approach, standard VMD with a finite energy-dependent width,\footnote{This produces an almost negligible effect for $V=\omega$, but guarantees sensible results for $V=\phi$.}
and a chiral Lagrangian treatment with light vector mesons from Refs.~\cite{Terschlusen1,Terschlusen2}, 
to data from Refs.~\cite{LeptonG,NA60,NA60new}. 
The dispersive approach leads to a significant enhancement of the transition form factor over the pure VMD result, 
which in turn results in an improved description of the data.
Part of this enhancement is even present if we use a simplified, VMD-inspired $\omega\to3\pi$ partial wave
$f_1(s)=a\,\Omega(s)$ inside the dispersion integral.
As Fig.~\ref{fig:f1} suggests, using the correct full $\omega\to3\pi$ P wave mainly leads to a further enhancement
for invariant masses of the lepton pair near and slightly above the two-pion threshold.
We note that using the slightly smaller sum-rule value for the normalization of the form factor
(instead of the one determined from the experimental $\omega\to\pi^0\gamma$ width) would 
further enhance $|F_{\omega\pi^0}(s)-1|^2$ by 5--10\%, albeit at the expense of a significantly enlarged uncertainty.
However, we also find that our analysis cannot account for the steep rise towards the end of the decay region, 
which is somewhat better described by the calculation in Refs.~\cite{Terschlusen1,Terschlusen2}. 
The size of the discrepancy for large invariant masses is surprising (note that the form factor in Fig.~\ref{fig:TFF} 
is shown on a logarithmic scale), in particular given the level of agreement found in the sum rule for the $\omega\to\pi^0\gamma$
branching ratio that should converge rather worse.  
Within the dispersive framework it is therefore hard to think of a plausible explanation for such a steep rise.
We note that in contrast to $\phi\to3\pi$, 
we have not yet been able to test the $\omega\to3\pi$ decay amplitude against experimental precision studies of the Dalitz plot, 
so a remaining deficit in our input for $f_1(s)$ cannot rigorously be excluded.
Still, given the analogy to the $\phi\to3\pi$ study in Ref.~\cite{V3pi},
it is implausible that this can account for the size of the difference.

The transition form factors are often characterized by their slope at $s=0$:
\begin{equation}
b_{V\pi^0} = \frac{dF_{V\pi^0}(s)}{ds}\bigg|_{s=0} ~.
\end{equation}
We quote this slope in units of $M_\rho^{-2}$ (where we use $M_\rho=775.5$~MeV), such that VMD
suggests $b_{\omega\pi^0}=1\,M_\rho^{-2}$.  Our dispersive analysis yields
\begin{equation}
b_{\omega\pi^0}= (1.41\ldots 1.45)\, M_\rho^{-2} ~, \label{eq:Nbompi}
\end{equation}
therefore a significant enhancement with respect to the VMD value, yet not as large
as the theoretical value found in Refs.~\cite{Terschlusen1,Terschlusen2}, $b_{\omega\pi^0}\approx 2\,M_\rho^{-2}$, 
and significantly smaller than the experimental determinations
$b_{\omega\pi^0}= (3.72\pm0.10\pm0.03)\,M_\rho^{-2}$~\cite{NA60},
$b_{\omega\pi^0}= (3.73\pm0.04\pm0.05)\,M_\rho^{-2}$~\cite{NA60new}. We note, however, that the latter experimental 
extractions are in principle model-dependent, as they rely on a monopole parameterization.

In order to improve on the comparison of our form-factor description to the data, 
one may think of subtracting Eq.~\eqref{eq:TFFdisp} once more and treating the additional
subtraction constant as a free parameter, at the expense of spoiling the high-energy behavior
of the transition form factor.  The difference between the once- and twice-subtracted representation
amounts to an additive term $\Delta b_{\omega\pi^0}\times s$, and it is rather obvious that this term
cannot account for the strong \emph{curvature} in the form factor at higher energies, such that the overall
picture is not drastically improved.  Furthermore, the value given in Eq.~\eqref{eq:Nbompi}
amounts to a value for the slope given by a sum rule, which would be expected to converge much better than the one 
for $f_{\omega\pi^0}(0)$ in Eq.~\eqref{eq:TFFsumrule}, yet it yields a result ostensibly off by a large factor.

The differential decay width for $\omega\to\pi^0\mu^+\mu^-$ (top, right panel in Fig.~\ref{fig:TFF}) 
is calculated according to Eq.~\eqref{eq:Decayrate}.\footnote{The normalization of the VMD prediction is obtained 
from the experimental $\omega\to\pi^0\gamma$ partial width, similar to the dispersive calculation. 
We refrain from displaying errors on the VMD calculation thus induced, 
since it merely serves illustrative purposes.} 
We observe that the values of the form factor close to the end of the decay region are actually strongly suppressed
by phase space. From that vantage point the situation does not look as dire as when the form factor is 
considered directly; however, due to the smallness of the errors of those values our solution still
deviates by several $\sigma$. The integration of the spectrum yields
\begin{equation}
 \BR(\omega\to\pi^0\mu^+\mu^-)=(0.94\ldots 1.00)\times 10^{-4}~,
\end{equation}
which agrees with the experimental average 
$
 \BR^{\rm exp}(\omega\to\pi^0\mu^+\mu^-)= (1.3\pm 0.4)\times 10^{-4}
$~\cite{PDG} within errors. 
This is not surprising: as the largest deviations from the experimental form factor
are strongly suppressed by phase space, they do not have a 
large influence on the partial width.

We also display the $\omega\to\pi^0e^+e^-$ differential decay width (bottom, right panel in Fig.~\ref{fig:TFF}), 
which has not been measured yet. 
Phase space combined with the $1/s$ behavior of the virtual photon lead to a strong enhancement near threshold
and a variation of the spectrum over many orders of magnitude;
we therefore only display the full dispersive result, since it is almost indistinguishable from $f_1(s)=a\,\Omega(s)$
on this scale. 
For a better comparison to $\omega\to\pi^0\mu^+\mu^-$, we also show this spectrum restricted to energies
$\sqrt{s}\geq 2m_\mu$.  As can be seen in Fig.~\ref{fig:TFF}, both leptonic final states yield very similar 
amounts of events in this energy range, where form-factor effects (deviations from pure QED) are felt most strongly.
The integrated spectrum for $\omega\to\pi^0e^+e^-$ yields
\begin{equation}
 \BR(\omega\to\pi^0e^+e^-)=(7.6\ldots 8.1)\times 10^{-4}~, \label{eq:BRompiee}
\end{equation}
where the uncertainty is dominated by the normalization given by $\BR(\omega\to\pi^0\gamma)$---the 
$s$-dependent $e^+e^-$ spectrum
is largely given by pure QED.  Equation~\eqref{eq:BRompiee} is
in perfect agreement with the experimental value
$
 \BR^{\rm exp}(\omega\to\pi^0e^+e^-)=(7.7\pm 0.6)\times 10^{-4}
$
within uncertainties.

\subsection{\boldmath{$\phi\to\pi^0\ell^+\ell^-$}}

\begin{figure*}[t]
\centering
\centerline{
 \parbox{0.51\linewidth}{\includegraphics[clip=true,height=11.4cm]{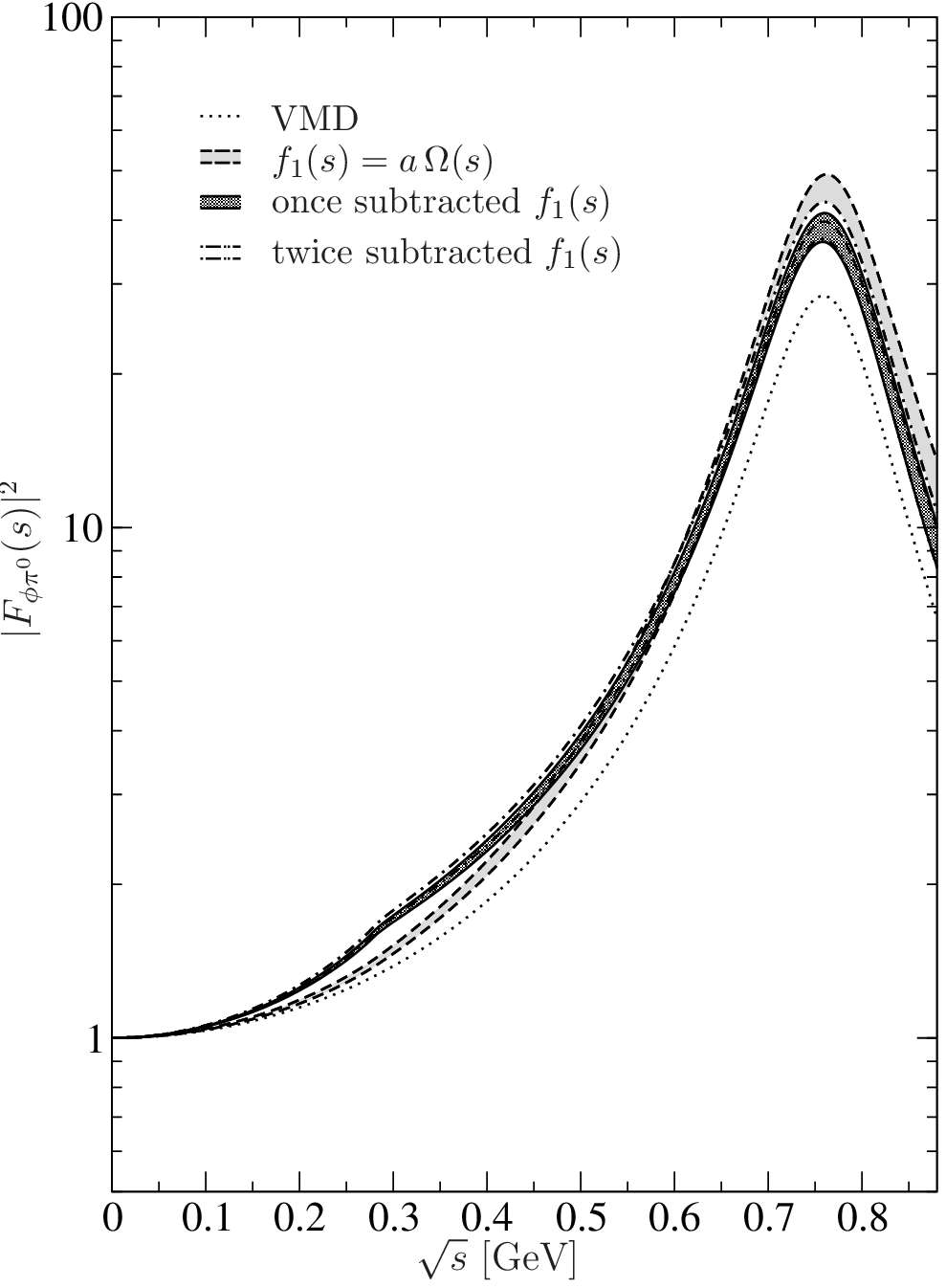}} \hfill
 \parbox{0.49\linewidth}{\vspace*{-2mm}\includegraphics[clip=true,height=5.4cm]{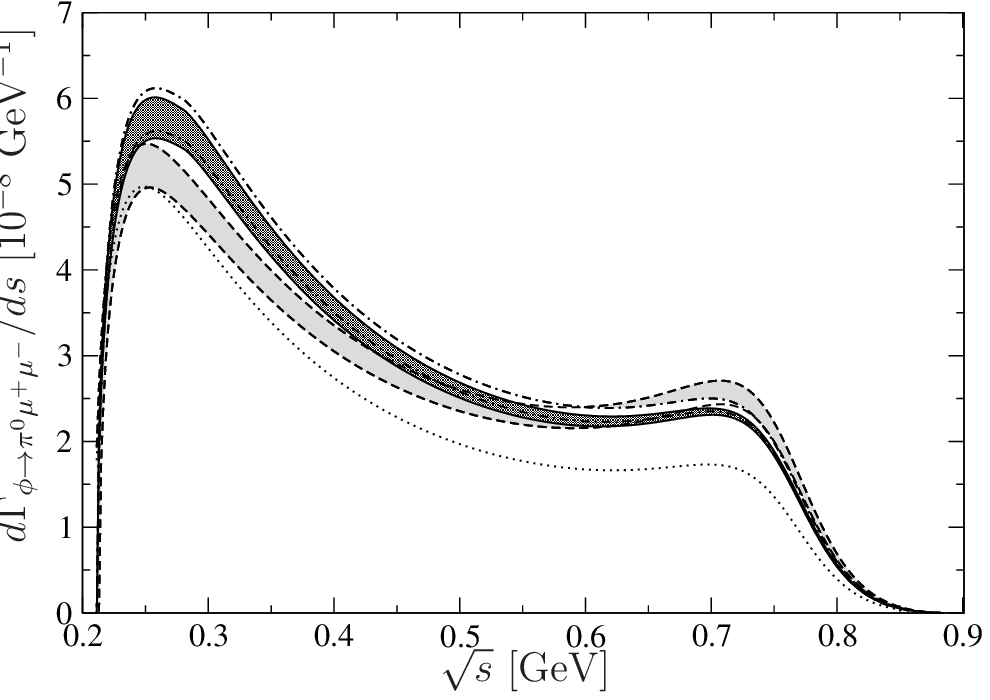} \\[1mm] \hspace*{-2.5mm}
                         \includegraphics[clip=true,height=5.55cm]{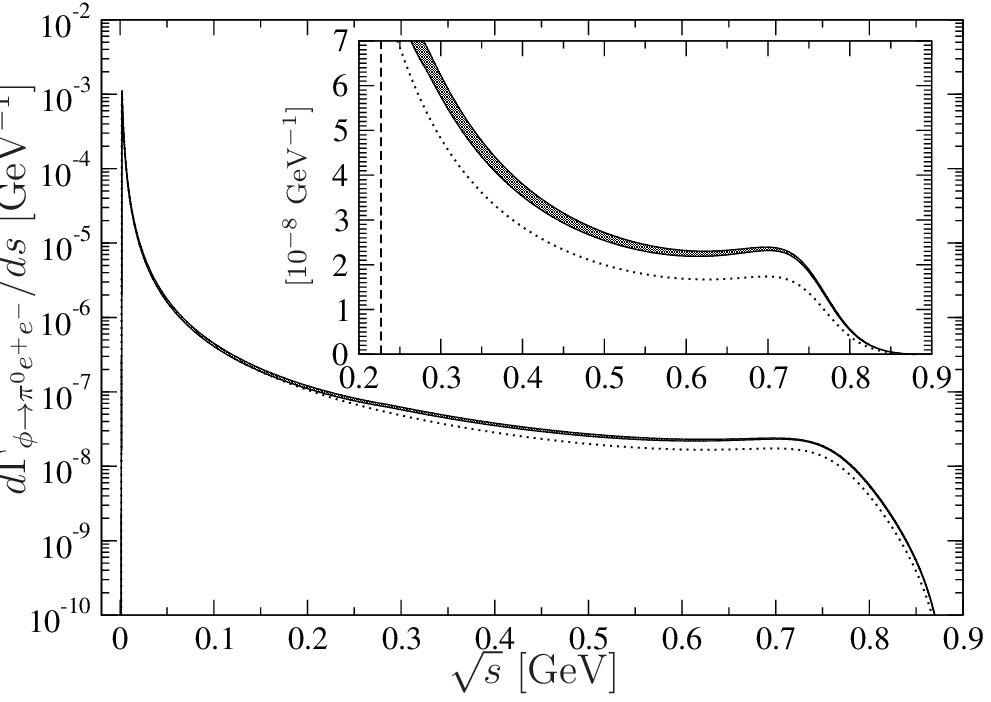}}}
\vspace*{-0.1cm}
\caption{Left panel: Numerical results for the $\phi\to\pi^0\gamma^*$ transition form factor.
Top, right panel: Differential $\phi\to\pi^0\mu^+\mu^-$ decay width. 
Bottom, right panel: Differential $\phi\to\pi^0e^+e^-$ decay width. 
We show pure VMD (dotted line), 
the dispersive solution for $f_1(s)=a\,\Omega(s)$ (gray shaded curve with dashed borders), 
and the full dispersive solution with one subtraction (black hatched curve with solid borders) 
and two subtractions (white shaded curve with dashed-dotted borders) in the $\phi\to3\pi$ partial wave. 
For $\phi\to\pi^0e^+e^-$ we only show the once-subtracted dispersive solution, since again neither
the Omn\`es solution nor the twice-subtracted one is visibly distinguishable 
from the once-subtracted result on the scale shown. The inset magnifies the region above the two-muon threshold 
(vertical dashed line).}
 \label{fig:PhiTFF}
\end{figure*}

Next we consider $V=\phi$. The results are displayed in Fig.~\ref{fig:PhiTFF} 
for the absolute value squared of the $\phi\to\pi^0\gamma^*$ transition form factor (left panel) 
and the differential $\phi\to\pi^0\ell^+\ell^-$ decay widths (right panel). 
There are no experimental data in any of the decay channels yet. 
For the $\phi\to\pi^0\gamma^*$ transition form factor we observe a similar behavior 
as for the $\omega$: 
in the full decay region, 
the form factor calculated with the dispersive approach is enhanced compared to the pure VMD result; 
in addition, we observe the two-pion-threshold enhancement of the full dispersive result with respect to 
$f_1(s)=a\,\Omega(s)$.  Due to the strong rise of the full solution for $f_1(s)$ towards this threshold, 
see Fig.~\ref{fig:f1},
the corresponding $F_{\phi\pi^0}(s)$ almost approaches a cusp-like behavior at $s=4\mpc^2$.
Since the $\phi$ as opposed to the $\omega$ transition form factor encompasses the $\rho$ resonance region, 
we can also observe that the full solution for $f_1(s)$ slightly reduces the height of the resonance peak with respect to the 
simplified assumption $f_1(s)=a\,\Omega(s)$, which agrees with our observations in Ref.~\cite{V3pi}. 
We note that using a twice-subtracted dispersion relation for the partial-wave amplitude $f_1(s)$, 
with the additional subtraction constant fitted to the $\phi\to3\pi$ Dalitz plot of Ref.~\cite{KLOEphi}, does not change 
our results by all that much: the differences are smaller than the overall uncertainty in our transition-form-factor
prediction.  This corroborates our skepticism that an imperfect determination of the $\omega\to3\pi$ P wave is the likely source
of the discrepancy seen in the $\omega$ transition-form-factor data.

Again, we also quote the derivative of the form factor at $s=0$:
\begin{equation}
b_{\phi\pi^0}= (1.52\ldots 1.61)\, M_\rho^{-2} ~,
\end{equation}
which is still somewhat larger than $b_{\omega\pi^0}$, see Eq.~\eqref{eq:Nbompi}, but again not nearly as large as
the slopes found experimentally in other vector-meson conversion decays.

The observations above concerning the differences of the various theoretical predictions
translate directly to the $\phi\to\pi^0\mu^+\mu^-$ differential decay spectrum (top, right panel of Fig.~\ref{fig:PhiTFF}).
We find that the $\rho$ resonance leaves a clear imprint on the spectrum, as one observes a second 
peak structure that counterbalances the drop-off of the phase-space factor. The integrated spectrum yields
\begin{align}
 \BR^\text{once}(\phi\to\pi^0\mu^+\mu^-) =(3.7\ldots 4.0) \times 10^{-6} ~, \nnnl
 \BR^\text{twice}(\phi\to\pi^0\mu^+\mu^-)=(3.8\ldots 4.1) \times 10^{-6} ~,
\end{align}
for the once- and twice-subtracted $\phi\to3\pi$ partial-wave amplitudes, respectively,
perfectly compatible within the error ranges.
There is currently no experimental measurement of the partial width to be compared with.

As for the corresponding $\omega$ decay,
the differential $\phi\to\pi^0 e^+e^-$ decay width is enhanced for small $s$ by several orders of magnitude;
for this reason, we only display the full dispersive solution based on the once-subtracted $\phi\to3\pi$
partial wave $f_1(s)$, the alternatives being indistinguishable on this scale.  
Again, an insert concentrates on energies above the two-muon threshold for better comparison 
of the expected event rates in both final states.
The results for the integrated spectra are
\begin{align}
 \BR^\text{once}(\phi\to\pi^0e^+e^-) &=(1.39\ldots 1.51)\times 10^{-5}~,\nnnl
 \BR^\text{twice}(\phi\to\pi^0e^+e^-)&=(1.40\ldots 1.53)\times 10^{-5}~,
\end{align}
for both of the full solutions, respectively. Compared with the experimental value of
$
 \BR^\text{exp}(\phi\to\pi^0e^+e^-)=(1.12\pm 0.28)\times 10^{-5}
$~\cite{PDG},
we find agreement within uncertainties.

We wish to emphasize the significance of an experimental investigation of the $\phi\to\pi^0\gamma^*$ transition form factor.
Deviations from the VMD picture now seem to be well established
in $\omega\to\pi^0\gamma^*$; strikingly enough, both this and the latest measurement of the transition form factor 
in $\phi\to\eta\gamma^*$~\cite{Zdebik}, when parameterized in terms of a monopole form factor, yield monopole mass parameters
significantly below the scale of the physical vector mesons, but (of course) too large to be accessible within the 
physical decay region.
This is different in $\phi\to\pi^0\gamma^*$: the $\rho$ resonance can be measured in this decay;
if there systematically is a steep form factor rise as seen in $\omega\to\pi^0\gamma^*$, 
mapping it out in full in $\phi\to\pi^0\gamma^*$ will help clarify its origin.
From the theoretical side, our dispersive analysis for this process
is based on a very precisely measured $\phi\to3\pi$ Dalitz plot, such that we are very confident about 
the reliability of our prediction.  
We thus strongly advocate an experimental analysis of the $\phi\to\pi^0\gamma^*$ form factor to the best possible precision.

\begin{figure}[t]
\centering
\includegraphics[width= \linewidth]{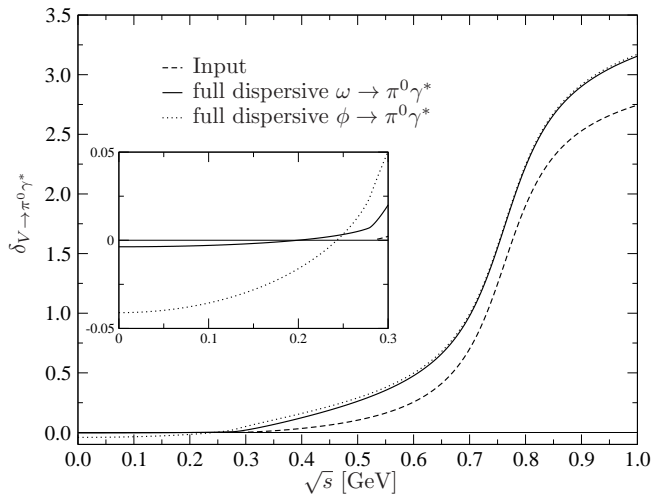}
\caption{Phase of the electromagnetic transition form factor for $\omega\to\pi^0\gamma^*$ (solid line) and $\phi\to\pi^0\gamma^*$ (dotted line) together with the input phase-shift parameterization (dashed line). 
The inset shows a magnification of the region below the two-pion threshold.}
 \label{fig:PhaseTFF}
\end{figure}
\begin{figure}
 \centering
\includegraphics[width= 0.75\linewidth,clip=true]{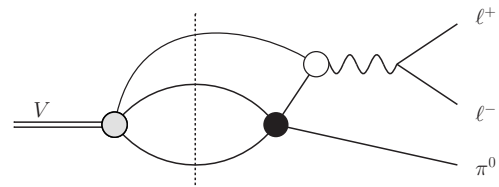} \\
\caption{Three-pion-cut contribution to the $V\to\pi^0\ell^+\ell^-$ transition vector form factor,
inducing an imaginary part also for $s<4\mpc^2$.}
\label{fig:TFFdisc-three}
\end{figure}

As a final illustration, we show the (experimentally unobservable) phases of the $V\to\pi^0\gamma^*$ transition form factors, 
both for $\omega$ and $\phi$, in Fig.~\ref{fig:PhaseTFF}. 
We calculate these from the \emph{unsubtracted} solution to the discontinuity equation~\eqref{eq:TFFdisc},
as we can only fix the modulus of the subtraction constant $f_{V\pi^0}(0)$ by means of
the $V\to\pi^0\gamma$ partial width, not its phase, which is nonvanishing due to the complex discontinuity of the 
$V\to3\pi$ partial-wave amplitude $f_1(s)$.
Only the phases of the full dispersive calculations, as compared to the $\pi\pi$ P-wave phase shift, are displayed. 
We refrain from showing the (small) error bands and fix the input in complete analogy to Fig.~\ref{fig:f1}. 
An additional consequence of the complex discontinuity of the $V\to3\pi$ partial-wave amplitude is 
that Watson's final-state theorem also does not apply to the transition form factors, 
and their phases are \emph{different} from $\delta(s)$: three-pion-cut effects, 
see Fig.~\ref{fig:TFFdisc-three},
produce nonvanishing imaginary parts/nonvanishing phases of the transition form factors also below the $\pi\pi$ threshold,
$s<4\mpc^2$. We observe the transition-form-factor phases to be significantly larger than $\delta(s)$ above the two-pion threshold,
and a tendency to small negative values below.

\section{Summary}\label{sec:summary}

In this article, we have analyzed the $V\to\pi^0\gamma^*$ electromagnetic transition form factors for $V=\omega$ and $V=\phi$ 
by means of a dispersive framework. It requires the corresponding P-wave projection of the $V\to\pi^+\pi^-\pi^0$ decay amplitudes, 
and the pion vector form factor as input, both of which depend on the pion--pion P-wave
scattering phase shift as input and are otherwise predictions up to a subtraction constant determining the overall
normalization of the $V\to3\pi$ amplitudes.
The consistent treatment of crossed-channel effects in the $V\to3\pi$ partial-wave amplitudes by
incorporating three-particle cuts leads to a nontrivial analytic structure for the transition form factors,
in particular its phase does not follow the $\pi\pi$ P-wave phase.

We have calculated the real-photon $V\to\pi^0\gamma$ branching ratios using a sum rule,
which yields good agreement with the experimental $\phi\to\pi^0\gamma$ branching ratio and indicates
that the sum rule for $\omega\to\pi^0\gamma$ (which is much more precisely determined experimentally)
is saturated roughly at the 90\%--95\% level by two-pion intermediate states.
To lessen the dependence on medium-to-high-energy input, we have oversubtracted the dispersion relation 
for the form factors and used the real-photon partial widths as input for the subtraction constant.
We found that this approach leads to an enhancement compared to a pure VMD calculation and thus to an 
improved description of experimental data from NA60 for the $\omega\to\pi^0\mu^+\mu^-$ channel. 
Three-pion effects in particular lead to an enhancement in the two-pion-threshold region.

We are unable to solve the puzzle of the steep rise in the $\omega\to\pi^0\gamma^*$ transition form factor data 
close to the end of the decay region. 
In order to try to better understand the physical mechanism behind this enhancement, 
we strongly advocate a measurement of the $\phi\to\pi^0\gamma^*$ transition form factor: 
the fact that the physical region of the decay goes beyond that of the 
corresponding $\omega$ decay and incorporates the $\rho$ resonance peak suggests that it should give some clues 
about the nature of this rise. 

While our predictions for branching ratios of the various $V\to\pi^0\ell^+\ell^-$ channels  are in good agreement 
with experimental determinations, data on decay spectra only exists for $\omega\to\pi^0\mu^+\mu^-$.
It would certainly be helpful, especially in light of
a theoretical analysis of contributions to light-by-light scattering, 
if precision data for additional channels could be obtained~\cite{WASA}.

\begin{acknowledgments}
We are grateful to S.~Leupold for extensive, very useful discussions and a critical reading of this manuscript.
We would like to thank S.~Damjanovic and G.~Usai for providing us with the data of Refs.~\cite{NA60,NA60new},
and G.~Colangelo for the preliminary $\pi\pi$ phase-shift results of Ref.~\cite{Bern}.
Partial financial support by
the DFG (SFB/TR 16, ``Subnuclear Structure of Matter''),
by the project ``Study of Strongly Interacting Matter'' (HadronPhysics3, Grant Agreement No.~283286) 
under the 7th Framework Program of the EU,
and by the Bonn--Cologne Graduate School of Physics and Astronomy
is gratefully acknowledged.
\end{acknowledgments}

\bigskip

\appendix

\section{Pion vector form factor representation including higher resonances}\label{app:resFF}

To account for the effects of higher resonances in the pion form factor, at least in the elastic approximation,
we use the analytic representation~\cite{Roig}
{\allowdisplaybreaks
\begin{align}
&F_\pi^V(s) 
= \frac{M_\rho^2+s(\gamma\, e^{i\phi_1}+\delta\, e^{i\phi_2})}{M_\rho^2-s-iM_\rho\Gamma_\rho(s)} 
\exp\bigg\{\!-\frac{s A_\pi(s)}{96\pi^2\Fpi^2}\bigg\} \nnnl
& - \frac{\gamma\, s\, e^{i\phi_1}}{\mrr^2-s-i\mrr\grr(s)} 
\exp\bigg\{\!-\frac{s\grr A_\pi(s)}{\pi \mrr^3\sigma_\pi^3(\mrr^2)}\bigg\} \nnnl
& - \frac{\delta\, s\, e^{i\phi_2}}{\mrrr^2-s-i\mrrr\grrr(s)} 
\exp\bigg\{\!-\frac{s\grrr A_\pi(s)}{\pi \mrrr^3\sigma_\pi^3(\mrrr^2)}\bigg\} \,,
\label{eq:roigFF}
\end{align}}\noindent
where
\begin{align}
A_\pi(s) &= \log\frac{M_\pi^2}{M_\rho^2} + \frac{8M_\pi^2}{s} - \frac{5}{3} 
+ \sigma_\pi^3(s)\log\frac{1+\sigma_\pi(s)}{1-\sigma_\pi(s)} \,, \nnnl
\Gamma_\rho(s) &= \frac{M_\rho s}{96\pi\Fpi^2} \sigma_\pi^3(s) \,, \quad \sigma_\pi(s) = \sqrt{1-\frac{4\mpc^2}{s}} \,,\nnnl
\Gamma_{\rho',\rho''}(s) & = \frac{M_{\rho',\rho''}}{\sqrt{s}}
\bigg(\frac{s-4\mpc^2}{M_{\rho',\rho''}^2-4\mpc^2}\bigg)^{3/2}
\Gamma_{\rho',\rho''} \,,
\end{align}
and $\Fpi$ is the pion decay constant.
We fit the various masses, widths, and coupling constants to the 
$\tau^-\to\pi^-\pi^0\nu_\tau$ data of Ref.~\cite{Belle} (thus eschewing the additional complication
of $\rho$--$\omega$ mixing).
Note that we have omitted the kaon loop contributions  to $\Gamma_\rho(s)$, so that our fit values are slightly 
different as compared to Ref.~\cite{Roig}. We obtain
\begin{align}
 M_{\rho'} &= 1.44 \pm 0.01~\text{GeV}~, & \Gamma_{\rho'} &= 0.34 \pm 0.03~\text{GeV}~,\nnnl
 \gamma &= 0.097 \pm 0.009~, & \phi_1 &= 0.5 \pm 0.2~,\nnnl 
 M_{\rho''} &= 1.71 \pm 0.05~\text{GeV}~, & \Gamma_{\rho''} &= 0.13 \pm 0.03~\text{GeV}~,\nnnl
 \delta  &= -0.02 \pm 0.02~, & \phi_2 &= 1.1 \pm 0.6~.
\end{align}

\end{document}